\documentclass[11pt,twoside]{article}
\usepackage{asp2010}
\usepackage{rotating}

\resetcounters

\begin{document}

\title{Ultracool Dwarf Science from Widefield Multi-Epoch Surveys}
\author{N.R. Deacon$^1$, D.J. Pinfield$^2$, P.W. Lucas$^2$, Michael C. Liu$^1$, M.S.} \author{Bessell$^3$, B. Burningham$^2$, M.C. Cushing$^4$, A.C. Day-Jones$^5$,} 
\author{S. Dhital$^6$, N.M. Law$^7$, A.K. Mainzer$^4$ and Z.H. Zhang$^2$}
\affil{$^1$Institute for Astronomy, University of Hawai`i, 2680 Woodlawn Drive, Honolulu, Hawai`i, 96822-1839, USA}
\affil{$^2$Centre for Astrophysics Research, University of Hertfordshire, Hatfield, AL10 9AB, UK}
\affil{$^3$Research School of Astronomy and Astrophysics, Mount Stromlo Observatory, Cotter Rd,  ACT 2611, Australia}
\affil{$^4$Jet Propulsion Laboratory, M/S 264-765, 4800 Oak Grove Drive, Pasadena, CA 91109, USA}
\affil{$^5$Universidad de Chile, Camino El Observatorio 1515, Las Condes, Santiago, Chile}
\affil{$^6$Department of Physics \& Astronomy, Vanderbilt University, Nashville, TN 37235, USA}
\affil{$^7$Dunlap Institute for Astronomy and Astrophysics, University of Toronto, 50 St. George Street Room 101, Toronto, Ontario, M5S 3H4, Canada}

\begin{abstract}
Widefield surveys have always provided a rich hunting ground for the coolest stars and brown dwarfs. The single epoch surveys at the beginning of this century greatly expanded the parameter space for ultracool dwarfs. Here we outline the science possible from new multi-epoch surveys which add extra depth and open the time domain to study. 
\end{abstract}
\section{Introduction}
Widefield multi-epoch surveys have been a crucial discovery tool for ultracool dwarfs. Photographic plate data was used by \cite{LHS} to identify nearby cool stars by their proper motion. As technology evolved these plates were digitised (\citealt{Hambly2001}, \citealt{Monet2003}) leading to more discoveries of cool nearby stars.

The study of brown dwarfs did not begin with widefield digital infrared and red optical surveys, but datasets such as 2MASS (\citealt{Skrutskie2006}) and SDSS (\citealt{York2000}) massively expanded the sample of brown dwarfs. Works such as \cite{Chiu2006}, \cite{Reid2008} and \cite{Burgasser2004} provided large samples and objects from these surveys were used to define the new L and T spectral classes (\citealt{Kirkpatrick1999}, \citealt{Burgasser2006}).  

Now a new generation of widefield surveys are expanding into relatively unexplored regions, identifying extremely cool objects, finding nearby brown dwarfs by their trigonometric parallax and opening the time domain to the search for variability. Here we outline the leading surveys currently in operation and those about to come online.   
\section{Current Surveys}
\subsection{The UKIRT Deep Infrared Sky Survey}

Of the five UKIDSS surveys (\citealt{Lawrence2007}), three are contributing significantly to
the study of substellar objects: the Galactic Clusters Survey (GCS),
the Large Area Survey (LAS) and the Galactic Plane Survey (GPS). The
first of these has is targeting 10 clusters, and totalling over 1400
sq degs to a depth of $K=18.7$. Four of these will have two epochs to
allow proper motion selection of members for robust IMF determination
(Pleiades, Alpha Per, Praesepe and Hyades).  Results already published
for several clusters are painting a picture of the substellar IMF that
is broadly consistent with previous studies. For an IMF of the form
$\psi(M)~\propto~M^{-\alpha}$, a value of $\alpha = 0.6$ appears to be
common (e.g. \citealt{Lodieu2007}; \citealt{Lodieu2009a}).

The LAS search for field brown dwarfs has now resulted in the
discovery of over 100 T dwarfs, making it the single largest
contributor to the known sample of these objects, has now provided the most
precise measurement of the space density of T6-T8 dwarfs to-date
(\citealt{Burningham2010}), revealing a significant dearth of these objects
compared to what would be expected given the initial mass function
observed in young clusters. The origin of this discrepancy is not
clear, but possibilities include a variant substellar
initial mass function over the lifetime of the Galaxy or problems with the evolutionary models used to
predict the space density.  In addition to studies of the IMF, the LAS
sample is producing additional results from searches for benchmark
systems (e.g. \citealt{Day-Jones2010}; \citealt{Zhang2010}) and the search
for halo T dwarfs (e.g. Murray et al. in prep.)

Despite the challenges associated with searching for T dwarfs in the
Galactic plane, the GPS is also beginning to deliver very cold
objects. The discovery of the extremely cool T10 dwarf UGPS J0722-0540
at a distance of just 4.1pc (\citealt{Lucas2010}), means that UKIDSS
T dwarfs are now on the threshold of probing the sub-500K regime,
where water clouds may start to be seen in the photosphere. Over the
next few years the synergy of the UKIDSS data set with the WISE survey
will likely reveal even cooler atmospheres, and possibly the first
observation of water clouds beyond the Solar System.
\subsection{VISTA public surveys}
The Visible and Infrared Survey Telescope for Astronomy (VISTA) is a UK 
built 4 metre class, near-infrared survey telescope located at the ESO site 
on Cerro Paranal, Chile. VISTA's wide field camera has sixteen 2048x2048 
Raytheon arrays with quantum efficiency $>$90\%. It produces tiles with 
dimensions of 1.5x1.0 degrees, compiled from 6 dithered pawprints. The 
image scale is 0.34 arcsec per pixel. During its first 5 years of operation 
VISTA is mainly dedicated to performing six public surveys, which began 
taking data in late 2009. These surveys include: UltraVISTA, the smallest 
of the surveys covering only 0.73 square degrees to a depth of J=26.6. 
VIDEO (Vista Deep Extragalactic observations), VMC (Vista Magellanic Cloud 
Survey), VVV (Vista Variables in the Via Lactea), a synoptic survey of the 
Galactic plane, VIKING (Vista Kilo-Degree Infrared Galaxy Survey) and the 
largest, VHS (Vista Hemisphere Survey), which will image the rest of the 
southern hemisphere.

The last two of these surveys will likely be the source of most brown dwarf 
discoveries from the VISTA surveys. With their large sky coverage in 
several filters, these surveys will probe deeper than any other 
near-infrared survey, for example VIKING alone will cover three times the 
volume of the UKIDSS Large Area Survey, providing data in the $Z$,$Y$,$J$,$H$ and $K_s$ 
filters. Both VIKING and VHS will have optical counterparts provided 
respectively from the ESO KIDS (Kilo degree survey) and DES (Dark Energy 
Survey), such that combining these surveys, and data from WISE, Skymapper and 
Pan-STARRS, will provide a powerful tool for the identification of brown 
dwarfs. There is the potential to increase the number of T dwarfs know by 
an order of magnitude. In addition this will allow the identification of 
many brown dwarfs that could be used as benchmark objects as well as 
objects that are cooler than those presently known. These discoveries will 
be able to test and help calibrate existing ultracool models and 
measurement of the mass function to new precision. However, a very 
efficient follow up strategy will be required to classify the large number 
of brown dwarf candidates, many of which will be challenging targets for 
ground-based spectroscopy.
\subsection{The WISE All-Sky Survey}
One of the two primary science objectives for the Wide-field Infrared Survey
Explorer (WISE) is to find the coldest brown dwarfs, which represent the
final link between the lowest mass stars and the giant planets in our own
solar system. WISE is a NASA Medium-class Explorer mission designed to
survey the entire sky in four infrared wavelengths, 3.4, 4.6, 12 and 22
microns (\citealt{Wright2010}; \citealt{fliu2008}; \citealt{Mainzer2005}). WISE
consists of a 40 cm telescope that images all four bands simultaneously
every 11 seconds. It covers nearly every part of the sky a minimum of eight
times, ensuring high source reliability, with more coverage at the ecliptic
poles. Astrometric errors are less than 0.5 arcsec with respect to 2MASS
(\citealt{Wright2010}). The preliminary estimated SNR=5 point source
sensitivity on the ecliptic is 0.08, 0.1, 0.8 and 5 mJy in the four bands
(assuming eight exposures per band; \citealt{Wright2010}). Sensitivity
improves away from the ecliptic due to denser coverage and lower zodiacal
background. WISE's two shortest wavelength bands, centered at 3.4 and 4.6 m
(W1 and W2, respectively), were specically designed to optimize sensitivity
to the coolest types of brown dwarfs (Kirkpatrick et al. in prep). In
particular, BDs cooler than $\sim$1500 K exhibit strong absorption due to the
nu-3 band of CH4 centered at 3.3 microns, with the onset of methane
absorption at this wavelength beginning at $\sim$1700 K (\citealt{Noll2000}). The
as-measured sensitivities of 0.08 and 0.1 mJy in W1 and W2 allow WISE to
detect a 300 K BD out to a distance of $\sim$8 pc, according to models by \cite{Marley2002} and \cite{Saumon2008}. While the W1 and W2 bands are
superficially similar to the Spitzer/IRAC bands 1 and 2 (which were also
designed to isolate cool BDs; \citealt{Fazio1998}), the WISE W1 band is wider
to improve discrimination between normal stars and ultra-cool BDs (\citealt{Wright2010}. This results in a systematically larger W1-W2 color compared to
[3.6]-[4.5]. Currently, only about two dozen objects with spectral types
later than T7 are known, and four objects are known to have type T9 or later
(\citealt{Warren2007}; \citealt{Delorme2008}, \citealt{Burningham2008}, and \citealt{Lucas2010}). New spectral indices for typing these late-type T dwarfs have
been developed as the CH4 absorption band depths may not continue to
increase in the NIR with temperatures lower than those of T8/T9 dwarfs
(\citealt{Burningham2010}), but spectral anchors have yet to be defined.  We
have recently reported the discovery of WISEPC J0458+64 (\citealt{Mainzer2010}) an object which is consistent with an extremely late T dwarf.  The
best-fitting model has an effective temperature of 600 K, log g=5.0,
[Fe/H]=0, and evidence for the presence of vertical mixing in its
atmosphere. As this remarkably cool object was found easily in some of the
first WISE data, WISE is likely to find many more similar objects, as well
as cooler ones, inevitably producing abundant candidates for the elusive Y
class brown dwarfs. Scaling from the Spitzer sample of 4.5 micron selected
ultra-cool BD candidates (\citealt{Eisenhardt2010}), we expect to find
hundreds of new ultra-cool brown dwarfs with WISE. We compute that for most
likely initial mass functions, WISE has a better than 50\% chance of
detecting a cool BD which may actually be closer to our Sun than Proxima
Centauri (\citealt{Wright2010}).
\subsection{Surveys with Pan-STARRS\,1}

The Pan-STARRS project (Panoramic Survey Telescope and Rapid Response
System; \citealt{Kaiser2002}), led by the University of Hawaii's Institute
for Astronomy, is developing a unique optical survey instrument
consisting of four co-aligned wide-field telescopes based in the
Hawaiian Island ({\tt http://panstarrs.ifa.hawaii.edu}). As a pathfinder
for the full system, the project has successfully completed a single
telescope system, called Pan-STARRS~1 (PS1), on the summit of Haleakala
on the Hawaiian island of Maui. PS1 is a full-scale version of one of
the four Pan-STARRS telescopes, with a 1.8-m primary mirror which images
a 7-degree$^2$ field-of-view on a 1.4-Gigapixel camera. The large
\'etendue of PS1 allows for rapid surveying of the observable sky,
obtaining $\approx$2500~sq. degs and 1.5~Tb of raw data each night. 

The PS1 3.5-year science survey mission officially began in May 2010 and
is carrying out a suite of pre-defined surveys, spanning the nearest
asteroids to the cosmological horizon. Of greatest interest to ultracool dwarf research is the Pan-STARRS\,1\,3$\pi$ survey, the
largest of the PS1 surveys, amounting to 56\% of the observing time. The
Pan-STARRS\,1\,3$\pi$ survey will monitor three quarters of the sky (30,000 sq. degs north of
declination $-$30$^{\circ}$) in $g$, $r$, $i$, $z$ and $y$ with six epochs
per filter spread over three years and two observations per epoch to
detect solar system objects. The survey goes to depths roughly similar
to SDSS in the bluest bands and is significantly more sensitive in the
far-red, with the novel $y$-band filter (0.95-1.03~\micron) providing
the greatest gains for ultracool dwarf science. The 5$\sigma$
single-epoch limits are $grizy = 23.2, 22.5, 22.2, 21.2, 19.8$~mag (AB
system).

The Pan-STARRS\,1\,3$\pi$ cadence will allow objects to be selected based on their
proper motions and parallaxes without a priori colour selection
(\citealt{Magnier2008}). This will allow a complete survey of the ultracool
dwarf population within 30~pc down to the mid/late T regime, as well as
discovery of low-mass objects with extreme properties that might
otherwise have gone undetected. The resulting catalog — unprecedented in its volume, sample size, and
completeness — will address several key areas, including: (1) the
luminosity function of nearby ultracool objects with the first
volume-limited sample of brown dwarfs; (2) rigorous tests of theoretical
models of brown dwarf atmospheres and evolution using densely populated
color-magnitude diagrams; and (3) exploring the physics of ultracool
objects over a wide range of metallicities and gravities.

As of this writing, Pan-STARRS\,1 is just completing its first pass of
the sky. With these first epoch data, we have been carrying out a proper
motion survey for nearby L~and T~dwarfs by combining them with 2MASS. The survey spans a $\approx$10~year baseline and is most sensitive to
objects with proper motions from $\approx$0.2--2.0\arcsec/yr. So far this has resulted in the discovery of a number of interesting
ultracool dwarfs, including bright T dwarfs not previously identified in
2MASS (Deacon et al., in prep.).

In addition, the Pan-STARRS\,1 Medium Deep Fields is repeatedly
observing a set of high-galactic latitude fields to search for
supernovae and other transients (100~sq. degs in total). This survey is
also suitable for the stellar/substellar variability studies, with the
potential to find around 300 low-mass eclipsing binaries, 10 brown-dwarf
EBs, and about a dozen transiting brown dwarfs around low-mass stars
within 100~pc (\citealt{Dupuy2009}).

\subsection{The Stromlo Southern Sky Survey}
SkyMapper is a 1.3m telescope with a 5.7 sq degree field of view covered with 32 2Kx4K E2V CCds that will carry out a 6-color, multi-epoch survey of the southern sky (including the galactic plane) - The Stromlo Southern Sky Survey.  We aim to provide star and galaxy photometry to better than 3\% global accuracy and astrometry to better than 50 mas.
The sampling will be 4 hr, 1 day, 1 week, 1month and 1 yr, although not in all filters.  It will take five years to complete the survey. Some time is reserved for non-survey work.
The photometric system of u (like Stromgren), v (like DDO38), griz (SDSS) is designed to maximize precision in the derivation of stellar astrophysical quantities. We also have available an Halpha filter for limited use.  We expect 6 epoch limiting magnitudes of 22.9, 22.9, 22.8, 22.8, 21.9, 21.2 in u,v,g,r,i,z, respectively. The data will be supplied to the community after science verification without a proprietary period. 
Young M dwarfs will be surveyed using Halpha, r and i. L and T dwarfs (high-z QSO contaminants) will be surveyed using i-z. VISTA photometry will also be cross-referenced and proper motions derived from our different epoch SkyMapper observations as well as through comparison with earlier epoch catalog positions. Follow-up spectroscopy will be carried out on the ANU 2.3m telescope and AAO 3.9m telescopes.
\subsection{The Palomar Transient Factory} 
The Palomar Transient Factory (PTF; \citealt{Law2009}) is a new fully-automated,
wide-field survey conducting a systematic exploration of the optical
transient sky. The transient survey is performed using a new 8.1
square degree camera installed on the 48 inch Samuel Oschin telescope
at Palomar Observatory; colors and light curves for detected
transients are obtained with the automated Palomar 60 inch telescope.
With an exposure of 60 s the survey reaches a depth of m$\sim$21.3 and
m$\sim$20.6 (5$\sigma$, median seeing). Cadences range between 10 minutes and 5
days, with a 3-day-cadence $r$-band supernova search making up the
largest fraction of the survey. The survey covers 8000 square degrees,
and fields are observed between tens and hundreds of times. PTF
provides automatic, real-time transient classification and follow-up,
as well as a database including every source detected in each frame.
As of Nov 22 2010, PTF had discovered and spectroscopically classified
910 supernovae. The wide-area, many epoch dataset is being used for a
wide variety of cool star science, including the activity / rotation /
age relation, proper motion surveys, and a search for transiting
planets around 100,000 M and L dwarfs.
\section{Future Surveys}
\subsection{The Large Synoptic Survey Telescope}
The Large Synoptic Sky Survey (LSST) will cover the 20000 square
degrees of the Southern sky, with $\sim$1000 total visits in the
$ugrizy$ bands over 10 years. The faintness limit will be $z\sim$ 23.3
mags for each visit and $z\sim$ 26.2 mags for the total survey (\citealt{LSSTbook}). LSST's depth, all-sky coverage, and 1000
epochs provide an large and unique dataset for studying the ultracool
dwarfs (UCDs; specifically of  M, L, T, and the hypothetical Y
spectral classes). 

\begin{figure}
\begin{center}
\includegraphics[scale=0.5 ]{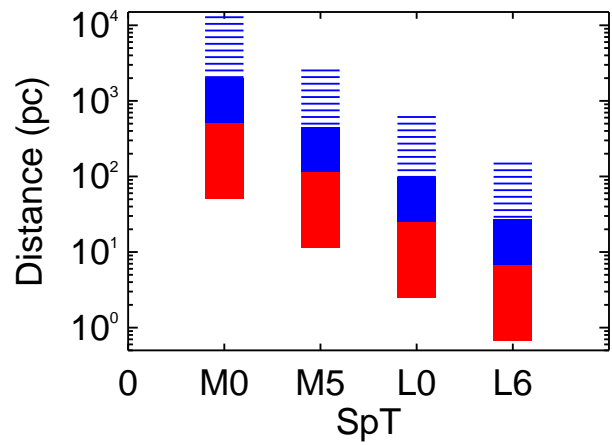} 
\end{center}
\caption{The distance at which stars of a given spectral type will
be detected in LSST. The red (lighter grey in black and white) strip shows the distance for the
brightness limits of $r=16$ and $r=21$ where proper motions greater
than  1.5~mas~yr$^-1$ and parallaxes greater than 4.5~mas will be
measured, with a 5=$sigma$ significance. The solid blue (darker grey in black and white) shows the
single epoch faintness limit, where proper motions greater than
7.5~mas$yr^-1$ and parallaxes greater than 22~mas will be
measured. The hashed blue shows the limits of the entire 10~year
survey, extending up to $r=28$. The depth as well as the proper motion
and parallax accuracies are unprecedented.}
\label{LSSTfig}
\end{figure}

LSST will allow for a census of the ultracool dwarfs (UCDs) in the solar
neighborhood, with M0, L6, and T9 dwarfs detectable to $\sim$10~kpc,
200 pc, and 25 pc, respectively. In terms of absolute numbers,
Galactic simulations predict$>$347,000 M, 35000 L, 2300 L, and $\sim$
18 Y dwarfs in the LSST fields (\citealt{LSSTbook}). In addition, we will have accurate proper
motions and parallaxes (see Figure~\ref{LSSTfig}). This will allow us not only to
measure the mass function, lumninosity function, and velocity
distribution of UCDs but also complete the census of nearby moving
groups and associations (also see chapter on Juvenile UCD session).

The Southern sky is very rich in large, nearby open clusters, like the
Orion Nebula, NGC 3532, IC 2602/IC 2391, and Blanco 1. Detecting and
characterizing brown dwarfs in these clusters, which have known ages, will
allow us to constrain the theoretical brown dwarf cooling curves
(e.g. \cite{Chabrier2000}). As most known brown dwarfs are in the
field with very uncertain ages, they cannot be used to constrain the
cooling curves. Similarly, LSST will detect a large of of UCD
eclipsing binaries that will serve as benchmarks to constrain our
stellar evolutionary models. Based on LSST data, 77 M and 2 L dwarfs
EBs will be characterized, with perhaps ten times as many UCD EB
candidates (J. Pepper et al. in prep.)
\section{Combining survey data}
As mentioned previously surveys often complement each other by providing information in different wavelengths or providing additional epochs for proper motion measurement. Works such as \cite{Burgasser2004} use infrared colours in combination with legacy optical data such as \cite{Monet2003} to select ultracool dwarfs while UKIDSS studies such as \cite{Burningham2010} make use of SDSS data. Examples of using widefield surveys in combination for proper motions include \cite{Sheppard2009} and \cite{Deacon2009a}. The new generation of surveys can also be used together and with legacy data increase their science yield.
\subsection{Ultracool Subdwarf Binaries Discovered from Large Area Surveys}
Ultracool subdwarfs (UCSDs; e.g., \citealt{Burgasser2009}) are metal- 
poor, halo counterpart of ultracool dwarfs. They have not been  
understood well both observationally and theoretically due to the  
limited number of known UCSDs, especially benchmarks (e.g., \citealt{Pinfield2006}). The number of known UCSDs is increasing benefit from  
current optical and NIR large area surveys (SDSS, \citealt{York2000};  
2MASS, \citealt{Skrutskie2006}; UKIDSS, \citealt{Lawrence2007}. This make  
it is possible to identified UCSDs in wide binary systems. I selected  
a sample of objects with proper motions large than 100 mas/yr from the  
SDSS and USNO-B proper motion catalog (\citealt{Munn2004}). Around one  
thousand M subdwarfs are confirmed with SDSS spectra. Start with this  
subdwarf sample, I identified two extreme UCSDs (esdM6+esdK2,  
esdM6+esdM1) and six M subdwarfs (including an sdM1+WD system) in  
binary systems based on their common proper motions (Zhang et al. in  
preparation). The esdM6+esdK2 binary is the most exciting system with  
very high proper motions (444$\pm$3mas/yr). Distances for both  
components are estimated, and both consistent with 180$\pm$30 parsec.  
It is one of the widest ultracool systems (16740$\pm$2790AU), With a  
separation of 93$\arcsec$ at this distance. It's galactic velocity  
($U=-325.3 km/s$; $V=-259.5 km/s$; $W=8.4 km/s$) consistent with halo  
population. This is the first halo subdwarf benchmark in wide binary  
system. We have obtained optical spectra of both components and will  
do further analysis on them. This research indicated that the wide  
cool subdwarf binary fraction is similar to that of ultracool dwarfs  
($^{>}_{\sim}$0.01; e.g., \citealt{Zhang2010}; \citealt{Faherty2010}). It  
indicates that a large number of benchmarks are potentially  
identifiable, sampling the broad range in mass, age and metallicity  
need to adequately calibrate both atmospheric and evolutionary models.

\section{Summary}
Widefield surveys will continue to play a leading role in the discovery and study of ultracool dwarfs. From the widefield infrared surveys with UKIDSS, VISTA and WISE, to multi-epoch optical surveys with Skymapper, Pan-STARRS, PTF and in the future LSST, such surveys will identify the coolest objects, large samples for population studies and many interesting variable sources. A summary of the surveys discussed in this paper can be found in Figure~2.

\acknowledgements The authors would like to thank the University of Washington for hosting Cool Stars 16, Local and Scientific Organising Committees and our SOC contact Andrew West.
\bibliography{deacon_n}
\clearpage
\begin{landscape}
\begin{figure}[!ht]
\smallskip
\begin{center}
\centering
{\footnotesize
\label{surveystab}
\caption{A summary of the surveys discussed in this paper.}
\begin{tabular}{lcccccl}
\hline
Survey&Filters&Depth(Single Epoch)&Area (sq.deg.)&Epochs&Dates&Public Release\\
\hline
UKIDSS LAS&$Y$,$J$,$H$,$K$&$Y$=20&3700&2(3000sq.deg.)&2005-2012&2014\\
Pan-STARRS\,1&$g$,$r$,$i$,$z$,$y$&$y$=19.8(AB, 5$\sigma$)&33000&6 per filter&2010-2013&2014\\
VISTA VHS&$J$,$H$, partial $Z$,$Y$,$K$&$J$=20.2(AB, 5$\sigma$)&19000&1&2010-2017&2011-\\
Stromlo SSS&$u$,$v$,$g$,$r$,$i$,$z$&$z$=20.6(AB, 5$\sigma$)&22000&6 per filter&2010-2015&After calibration\\
WISE&3.4, 4.6, 12, 22 $\mu$m&$\sigma$=0.1 mJy in 4.6$\mu$m&All Sky&1&2010&2011-2012\\
PTF&$g$,$R$&$r$=21&8000&300+&2009-&2012\\
LSST&$u$,$g$,$r$,$i$,$z$,$y$&$z$=23.3(AB)&20000&1000 total&2018-2028&2018-\\
\hline
\end{tabular}
}
\end{center}
\end{figure}
\end{landscape}
\normalsize

\end{document}